# Failure Of The Bell Locality Condition Over A Space Of Ideal Particles And Their Paths

Warren Leffler
Department of Mathematics,
Los Medanos College,
2700 East Leland Road Pittsburg,
CA 94565
wleffler@losmedanos.edu

ABSTRACT

We construct a space of ideal elements (particles and their paths) to analyze certain aspects of quantum physics. The particles are taken from a model of particle interaction first described by David Deutsch (based on a different but related framework, that of MWI), and the paths are based on Richard Feynman's path-integral formulation of quantum mechanics. By combining the two systems we develop a new approach to quantum mechanics that eliminates various quantum paradoxes.

## I. INTRODUCTION

In mathematics the use of ideal elements is a familiar and longstanding method with beneficial results. Some examples are points at infinity in projective geometry, Kummer's ideal numbers via ideals in certain rings, hyperreal numbers (which includes infinitely small numbers—"infinitesimals"—and their infinitely large multiplicative inverses) over the real numbers, and so forth. In this paper we show that ideal elements can play an important role in physics also, revealing insights into various aspects of the standard theory of quantum mechanics.

In what follows we use the term "standard quantum mechanics" (SQM) to refer to a mathematical representation of quantum phenomena derived (for configuration space) from canonical quantization. For both configuration and spin, SQM involves some form or other of Lebesgue square-integrable wavefunctions over a separable, Cauchy-complete Hilbert space. In this paper, however, we develop a new theoretical framework for quantum mechanics (the "space of all paths," SP). We do this by incorporating into the system certain kinds of ideal elements—namely, ideal particles and their paths.

Our theory, SP, is a straightforward merging of two separately formulated systems, one founded on David Deutsch's model of particle interaction based on Hugh Everett's MWI (Deutsch called our ideal particles, "shadow particles" [1]), and the other founded on Richard Feynman's path-integral approach to quantum mechanics [2]. Deutsch developed his model to illustrate, among other things, how MWI explains single-particle interference paradoxes, but MWI does not apply to multi-particle phenomena treated in Bell-type theorems [3, 4]. In contrast, in SP there is a single outcome for each quantum event, and this makes it, in particular, a suitable framework for examining Bell's theorem [5] for two entangled particles. In this connection, we will now discuss a fundamental idea, treated in a new way in this paper.

According to a widely-held view, "Bell's theorem proves that all hidden-variable theories based on the joint assumption of locality and realism are at variance with the predictions of quantum physics [6]." This is certainly true for SQM, because of Bell's obviously valid argument (and others similar to it) over a space that admits a measure, a probability distribution. However, the space of all continuous paths joining two points is nonmeasurable [7], although (because of Feynman's path-integral result) this framework, SP, replicates the experimental predictions of quantum mechanics. Moreover, it is easy to see rather quickly why Bell's argument breaks down in our SP framework.

Thus consider two space-like separated systems involving a pair of entangled particles that travel out in opposite directions from a source. Bell's universally accepted notion of locality for such a system is as follows [8, 9, 10]: Suppose that a measurement $\alpha$ is performed on one system, obtaining outcome $a$, and a measurement $\beta$ on the other system, with outcome $b$. Let $P(a,b|\alpha,\beta)$ be the probability for the joint outcome. Now suppose further that $\lambda$ (the so-called "hidden-variables" parameter [11]) represents a more complete description of the joint state of the two particles, one in which the joint probability factors into two independent local probabilities:

$$P(a,b|\alpha,\beta,\lambda) = P(a|\alpha,\lambda)P(b|\beta,\lambda). \qquad (1)$$

Therefore, according to Eq. (1), the probability of detection is now dependent only on $\lambda$ and the local measurement setting of $\alpha$ within the system involving that side; similarly for the oppositely located side and the setting $\beta$. This is sometimes called "the Bell factorizability condition" [12]. In a few short steps from Eq. (1), Bell derived an inequality satisfied by any local system (Bell's theorem) but violated by the quantum mechanical predictions. Hence the widely-held conclusion is that quantum mechanics requires a nonlocal theory.

But suppose that we take the hidden variables to be the set $\lambda$ of all possible paths $x(t)$ from the source to the detectors on each side of a two-particle experiment. (As Bell stated in Ref. [5]: "It is a matter of indifference in the following whether $\lambda$ denotes a single variable or a set, or even a set of functions, and whether the variables are discrete or continuous.") Yet, as noted above, this set is non-measurable—that is, there is no countably-additive and translation-invariant measure on the space of all continuous paths joining two points [Ref. 7]. Therefore, the very definition of locality represented by the factorizability property of Eq. (1) (which employs a probability distribution, a measure) is now meaningless over such a non-measurable space. The same holds of course for the subsequent steps in Bell's argument leading to his famous inequality—steps involving a translation-invariant, countably-additive probability distribution ($\rho(\lambda)$, in Bell's notation Ref. [5]).

Thus Bell's argument contains a tacitly assumed premise that fails in the SP framework: namely, measurability of the set $\lambda$ of all possible paths from the source event to the detectors.

We now illustrate how the above set of hidden variables represented by $\lambda$ enters into a typical two-particle physical setup. Thus, consider the Rarity-Tapster two-particle interferometer depicted on the left-half of Fig. 1, an experimental device that has long been used to produce correlations violating Bell's inequality [13].

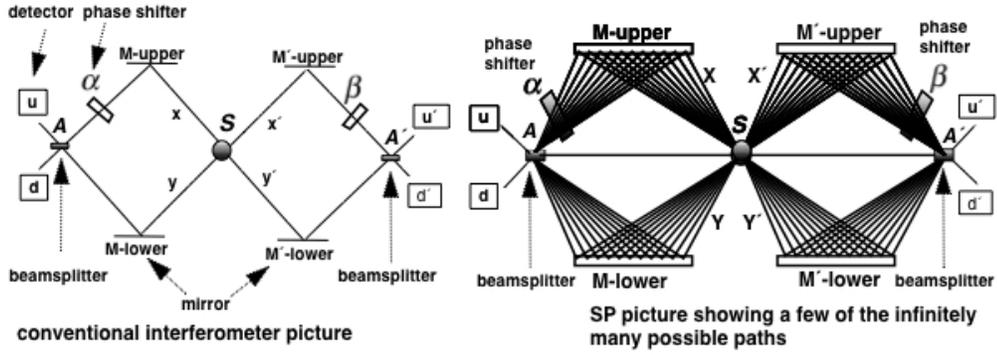

Fig. 1 At each run of the experiment, paired spinless particles travel first to floor-and -ceiling mirrors (M-M´lower and M-M´upper) on each side, from which they are then reflected to 50-50 beamsplitters at A and A´.

At each run of the experimental device depicted on the left-half of Fig. 1, two entangled spinless daughter particles (resulting from down conversion of a mother particle) travel out from a source to mirrors, from which they are reflected to beamsplitters on each side of the device, and then travel on to detectors [14]. The two particles take (by conservation of momentum) the paths x-y´ or y-x´, where measurements $\alpha$ and $\beta$ are performed on each side of the source. Hence, in the mathematical representation based on the SQM vector-space approach, one writes the quantum state of the pair of entangled particles as, say [15],

$$|\psi\rangle = (|x\rangle_L |y´\rangle_R + |y\rangle_L |x´\rangle_R)/\sqrt{2},$$

where the ket $|x\rangle_L$ denotes the particle on the left traveling along path x, and so forth, leading to the joint probability for correlations as:

$$\cos^2 \frac{(\alpha - \beta)}{2}.$$

But, as is well known, for correlations to occur in such a two-particle interferometer (unlike for a single-particle interferometer), there must be positional uncertainty at the source [16]. This generates an infinite, non-measurable set $\lambda$ of uncountably many possible paths from the source to the detectors, depicted in the right-half of Fig. 1.

In Sec. V (for the above kind of two-particle interferometer) we let $\lambda = L \cup R$ represent the set of infinitely many possible paths going from the source to the two separated systems represented by A and A´ at the beamsplitters in Fig. 1, with $L = X \cup Y$ and $R = X´ \cup Y´$. Then, by simple algebra involving correspondingly congruent paths between the sides, we show how to derive (summing over infinitely many paths) an analogue of the Bell factorizability condition for locality in Eq. (1) above:

$$\langle A | X \rangle_L \langle A | Y \rangle_L + \langle B | X´ \rangle_R \langle B | Y´ \rangle_R.$$

That is, although the underlying space is nonmeasurable, the amplitude leading to the probability for the correlations in the interferometer is just the sum of separate amplitudes from each side. This resembles the conventional condition, where (over a measurable space) the joint probability in Bell's definition of locality is the product of the separate probabilities on each side. Of course $\lambda = L \cup R$ comes from all possible paths within the experimental setup, but the set of all possible paths in any experimental arrangement is fixed at the outset and holds throughout all runs of the experiment. As implied in Postulate 1 below, what varies randomly at each run of the experiment are the two paths traveled by the pair of entangled tangible particles.

Regarding the relation between the Schrödinger's equation approach and the path-integral approach, Steven Weinberg notes:

> In his Ph.D. thesis, Richard Feynman (1918-1988) proposed a formalism, according to which the amplitude for a transition between one configuration of a set of particles at an initial time to another configuration at a final time is given by an integral over all the paths that particles can take in going from the initial to the final configuration. Feynman seems to have intended this path-integral formalism as an alternative to the usual formulation of quantum mechanics, but as later realized, it can be derived from the usual canonical formalism …. the path-integral formalism allows us to find the solution of the Schrödinger equation, without ever writing down the Schrödinger equation [17] .

The SP system is built around nonstandard elements, ideal particles and their paths. Yet, because of the well-known path-integral results, the two systems of SP and SQM are equivalent in their predictions ("SP" and "SQM" correspond in our terminology to what Weinberg calls the "path-integral formalism" and the "canonical formalism"). This guarantees immediately that SP replicates the quantum mechanical predictions of SQM.

Perhaps an instructive analogy is afforded by the mathematics of real analysis—where non-standard (infinitesimal) analysis over the hyperreal numbers is logically equivalent to the standard epsilon-delta approach over the real numbers. The two approaches provide different ways of arriving at the same mathematical result, but in a sense using radically different though conceptually related models of the real line [18]. Yet, although infinitesimals were important in the early history of calculus and related subjects, they were only placed on a rigorous foundation in the 1960s by Abraham Robinson, who did this using a construction from model theory in mathematical logic [19]. This validation came more than one hundred years after the rigorous development of the epsilon-delta approach by Augustin Cauchy and others in the nineteenth century. Mostly because of this head start (but also because of pedagogical inertia and the fact that the justification for nonstandard analysis relies on mathematical logic), the epsilon-delta approach is still taught almost exclusively in introductory calculus and even in advanced undergraduate analysis courses. Yet, as a consequence of Robinson's work, nonstandard analysis now enables us to picture rigorously each real number as surrounded by uncountably many ideal numbers (hyperreals) that differ by an infinitesimal amount from the given real number—a picture that corresponds to Leibniz's original view of the number line in calculus. As Robinson wrote [Ref. 18], "Gottfried Leibniz argued that the theory of infinitesimals implies the introduction of ideal numbers which might be infinitely small or infinitely large compared with the real numbers but which were *to possess the same properties as the latter*."

Although SQM and SP are equivalent in their predictions, they nevertheless differ in several important respects.

First, as we saw above, there is no measure on the space of all paths (unlike for the SQM Hilbert space of square-integrable wavefunctions), which has important consequences for Bell's theorem when it is considered within the context of SP. This nonmeasurability contrasts, for example, with what holds for a similar functional space, the Wiener space for Brownian motion [20]. But we note that *the essential requirement of a physical theory is certainly not that the underlying space must* *have a translation-invariant, countably-additive measure*. Instead, for quantum theory what is required is that we must be able to compute probabilities of quantum events, which we do in SP by summing (actually, taking the integral of) exponentiated-action terms over all possible paths to obtain the amplitude for the event, although the underlying space itself is nonmeasurable.

Second, SQM is opaque as to the mechanism that produces the observables, whereas in SP the hidden variables of ideal particles and their paths constitute a locally realistic framework underlying quantum phenomena, as we show in Sec. V. For example, a phase factor has no physical significance in SQM, since in SQM the expectation values of a self-adjoint operator are unchanged on multiplication by a complex number. On the other hand in the local realism of SP we postulate, for instance, that in traveling from point *a* to point *b* each particle in the tangible particle's "shadow stream" (which includes the tangible itself, Postulate I) has an initial (but unknown) phase in common with the other particles in the stream. The associated phase (unit vector, $\exp(iS[path]/\hbar)$) for each particle then rotates along to *b* according to the action over the particle's particular path. Because there are generally uncountably many paths between two points in the quantum space, this implies that at point *b* there is an enormous swarm of unit vectors pointing every which way. Convergence of the sum occurs because oppositely pointing vectors (corresponding to the "curled up" portion of the Cornu spiral in Fig. 2 below) cancel, leaving just those in the vicinity of the classical path to add constructively [21]. Thus the vector sum then carries (so we postulate) the new direction at point *b* of the tangible particle's unit vector, which (so we also postulate) is the same as that of all the others in the tangible's new shadow stream now generated at *b*, as the tangible travels to a further point *c*, and so forth. What is obvious is that this form of local realism in SP is clearly compatible with experiment—in particular for the single-particle and two-particle interferometer experiments that we discuss in Postulates 4 and 5 below and consider further in Sec. III and Sec. V.

Third, SQM is an epimorphic image of SP—under (typically) an infinitely-many-to-one mapping in which homotopy classes of paths in SP are mapped to basis vectors in SQM. (Recall that two paths with the same endpoints are homotopic when each can be continuously transformed into the other.) It is illustrated above with $\langle A | X \rangle_L$, etc., and is discussed further in Sec. III and Sec. V below. As a further example, consider a problem frequently treated from the SQM standpoint in elementary quantum mechanics—the so-called quantum particle-on-a-ring (roughly pictured by an electron moving on a conducting ring in, say, benzine). Here—using a Fourier expansion and the de Broglie wave length—one shows, for example, that there is a denumerably infinite set of eigenvalues for momentum. This can be treated also from the path integral perspective—which Lawrence Schulman did as a preliminary to developing a path integral for spin [22]. Thus suppose for the ring that we have a free tangible particle constrained to move on a circle and able to go any number of times back and forth along the circle. In SP we assume, as in Fig. 1 above, that the tangible is accompanied by ideal counterpart particles taking all the other possible paths. The underlying space is not simply connected, unlike Euclidean space, where Feynman originally developed his formulation. The homotopy classes of paths for the tangible's shadow stream correspond to winding numbers, *n*—that is, the number of times a particle travels past a given fixed point in the positive direction (counterclockwise) minus the number of times it passes the point in the minus direction (clockwise). We can thus map the homotopy classes of paths to basis vectors in SQM associated with winding numbers, so that the overall path-propagator corresponds to a denumerably infinite linear combination of such basis states.

Finally, we note that there is nothing special about the space of infinitely many possible paths in the above Rarity-Tapster two-particle interferometer. In fact (though apparently no one has considered it important until now) we claim that a similar configuration of infinitely many paths from the source to the detectors must be possible for any experimental arrangement involving two entangled particles, whether for a Rarity-Tapster two-particle interferometer, a Fransom-type interferometer [23], or for an experimental arrangement involving spin. (For a further instructive, but doubtless experimentally infeasible, gedanken experiment, one could also consider a pair of entangled ring particles as in the above example, where a source event imparts an initial momemtum kick that sends the shadow streams in opposite directions along a pair of rings that are somehow separated and moved apart following the source event.)

This quantum property of alternate possible paths, whether for single- or multi-particle interference, is related to the so-called *principle of indistinguishability*. This states that if we can, even in principle, distinguish the path that a tangible particle has taken, then there will no longer be interference [24]. Of course this rule in SQM is purely descriptive and explains nothing, although in SP it is explained by interference effects associated with counterpart ideal particles taking alternate possible paths (a picture of what underlies the collapse of the wavefunction in the Copehagen interpretation). Furthermore, we note that SP explains why any attempt to peer beneath the veil of the quantum event will cancel the interference—that is, because disruption of the tangible's shadow stream will alter the observable.

Apropos our counterargument to Bell's theorem, one could in principle (by checking the many experiments listed in the literature) verify that all two-particle Bell-type experiments implicitly involve infinitely many possible paths from the source event to the detectors. We haven't done that, however, because of this fact: *It takes only one counterexample to overturn a general concept or argument such as Bell's theorem, and our counterargument in Sec. V involving a Rarity-Tapster two-particle interferometer does that exhaustively and cleanly.*

Throughout this paper we will apply the well-known results of the path-integral approach in an essentially axiomatic way—simply citing the established facts when appropriate. What is novel in our approach derives from the argument above that Bell's condition [25, 26] for locality is meaningless in SP, and also the argument in Sec. V that SP is a local system. In Sec. III we analyze, in the SP framework, what's called "single-particle interference," showing how SP easily eliminates various paradoxes (a different way of accomplishing what Deutsch did using MWI in Ref. 1). In Sec. IV we describe a locality condition appropriate for the space of all paths. In Sec. V we give a counterexample in SP to Bell's theorem—a counterexample based on a standard Rarity-Tapster two-particle interferometer. We also sketch in Sec. V how SP applies to arguments involving more than two particles, such as GHZ [27].

## II. POSTULATES FOR THE SP FRAMEWORK

**Postulate 1 (Shadow stream):** With each separate quantum event corresponding to a "tangible" (ordinary) particle traveling between a pair of points (whether in configuration space or in spin space, $\mathbb{R}^3 \times SO(3)$), we associate a stream of ideal particles ("shadow" particles) traveling in a one-to-one fashion over all possible paths between the points. (As noted earlier, the terms "tangible" and "shadow" are from Deutsch [Ref. 1].) This unique set of paths and particles associated with each tangible particle traveling between the points we call "the shadow stream" generated by the tangible particle. We include the tangible particle in the shadow stream it generates, the tangible randomly taking one of the possible paths. As in Ref. 1, tangible particles only interact with shadow counterparts of the same type—tangible photons with shadow photons, tangible electrons with shadow electrons, and so forth.

Without loss of generality (the properties of the action functional extend easily to higher dimensions [28]), we will confine our discussion to $\mathbb{R}^2$. In particular with each path in the stream joining two points we associate a unit vector in the complex plane, exp($iS$[*path*]/$\hbar$), the "exponentiated-action" term over the path. The vector is assumed to be pointing in a certain, but unknown, direction at the initial point, leading also to a fixed direction at the final point, as in the Cornu spiral of Fig. 2 (thus, as noted earlier, the direction has physical significance, unlike in SQM).

**Postulate 2 (Amplitude generated by the sum of unit vectors):** The amplitude for a shadow stream associated with a tangible particle traveling between two points *a* and *b* is proportional to the sum of exponentiated-action terms over all possible paths between the points.

$$\sum_{\substack{\text{all paths } x(t) \\ \text{from } a \text{ to } b}} \exp(iS[x(t)]/\hbar).$$

Here $\hbar$ is Planck's constant $h$ divided by $2\pi$, "path" is the path of the particle, and $S$[path] is called the "action" over the path. As is well known, the exponentiated action acts like a "clock-pointer" that begins "ticking" at a certain rate, depending on the action—in our case, for each particle in the stream—as the particle proceeds along its path (of course, since the value of $\hbar$ is quite small, the ticking is at a very rapid rate). As usual, we take the probability of the corresponding quantum event to be proportional to the absolute square of the amplitude. Although this probability function is not translation-invariant and countably-additive in SP (and so, as we have been emphasizing, cannot be employed in carrying out Bell's argument), it nevertheless yields outcomes equivalent to those of standard quantum mechanics.

The action functional $S$ in coordinate space is based on the classical Lagrangian (integral over the difference of kinetic and potential energy). In spin space the action is the sum at each instant associated with a classical "top" action and a magnetic-field action along a rotational path in $\mathbb{R}^3 \times SO(3)$ [29]. For coordinate space, the action-functional approach has been known since the eighteenth century—from the work of Euler and Lagrange—to be equivalent to Newtonian mechanics in representing the motion of systems with algebraically described constraints [30].

In our system the classical action over each path joining two points pertains to a classical dynamic, with each quantum event being a sum over all such paths. Thus, if one is so inclined, it is possible to view a path and the ideal particle traveling it as belonging to something akin to a "world" in the sense of MWI, the tangible's quantum observable being a sum of exponentiated-action terms over all such worlds. But as pointed out earlier, in SP there is a single outcome for each quantum event, a crucial difference with MWI and the model that Deutsch based on MWI.

Now there was a time when the path integral (the sum over all paths) was felt to lack mathematical rigor, but that has greatly changed in the seventy years since Feynman discovered it, as the above quote from Weinberg illustrates. There are presently many approaches to the path integral, and its experimental equivalence to the standard framework has been shown over a wide class of configurations and potentials, for both coordinate space and spin space, with no contrary results [31]. (This includes a closed-form path-integral implementation of the experimental setup that we analyze as a counterexample to Bell's theorem in Sec. V—one involving a standard Rarity-Tapster two-particle interferometer.) Moreover, the path integral has become nearly indispensable in quantum field theory. Also, one should keep in mind that Bell's simple argument is clearly valid in a measurable space, with the consequence that correlations over widely separated regions then occur in a way that is utterly mysterious. As John Wheeler and Martin Gardner put it, the particles would have to "remain connected, even though light years apart, by a nonlocal sub-quantum level that no one understands ...." [32].

Following Feynman [33], we can graphically represent the sum-vector over the shadow stream (for the event illustrated by a free particle traveling from point $x_1$ to a mirror and then to $x_2$) by the "long" arrow in the Cornu spiral in Fig. 2 below. (As noted above, for simplicity, we will without loss of generality confine our discussion to $\mathbb{R}^2$. Also, although they are not depicted here, there are infinitely many paths that travel from the initial point to the final point without being reflected from the mirror. ).

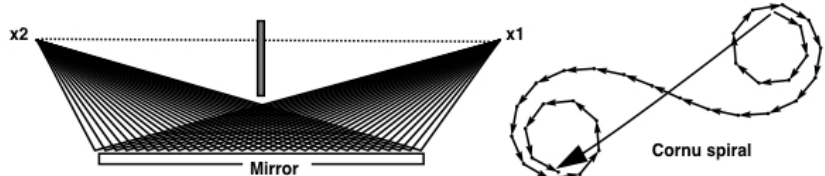

Fig. 2 Here we imagine a tangible particle, say a photon, accompanied by a stream of shadow photons. Each particle in the stream travels from x1 to a mirror, where it is then reflected up to point x2. The path of each such particle is associated with the exponentiated action, exp(iS[path])/ h-bar), over its path. This can be viewed as a "clockpointer" whose hand rotates at the classical frequency of the light. The rotation starts when the photon is emitted, and stops when the photon arrives at x2. For each particle when it reaches x2, the final position of its clockpointer produces a unit vector for that path. When the unit vectors are added head-to-tail they produce a vector sum, the "long" arrow in the Cornu spiral. The clockpointers associated with particles traveling paths of roughly equal time end up pointing in roughly the same direction, adding up constructively in the vector sum, while those in the "curled-up" portions tend to cancel.

The amplitude is approximated by the sum-vector in the Cornu spiral. Unit vectors $e^{iS[\text{path}]/\hbar}$ associated with the paths that take longer times for a particle to travel from $x_1$ to point $x_2$ in the figure tend to curl up in the spiral when added head-to-tail, canceling each other in the sum (pointing in opposite directions). On the other hand, vectors over paths of similar lengths tend to add up constructively in the sum-vector, the central portion contributing significantly to the length of the long vector. The absolute square of the magnitude of the long vector is proportional to the probability that the tangible particle in the stream pursues the so-called path of least time.

**Postulate 3** For *two non-interacting tangible particles* the composite amplitude of the two streams generated by the tangibles is *the product of the separate amplitudes*. (We are stating this postulate for convenience; it actually follows by an application of the rule, "The amplitude that one particle will do one thing and the other one do something else is the product of the two amplitudes that the two particles would do the two things separately." [34])

Now we will state two postulates that deal with 50-50 beamsplitters. The beamsplitter postulates are designed to duplicate the bivalent functions that determine which path a classical object pursues when it encounters a fork along a path in the toy model of Sec. IV. These postulates are intuitively obvious and consistent with the experimental outcomes of a standard interferometer, although they are based on local realism (again, we pointed out in Sec. I, Bell's theorem cannot rule out local realism in this context).

**Postulate 4 (Beamsplitter):** In SQM, there is no explanation for the quantum observable that occurs when a tangible particle encounters a beamsplitter, just a description of its behavior [35, 36]. As Feynman put it, "I am not going to explain how the photons actually 'decide' whether to bounce back or go through; that is not known. (Probably the question has no meaning.) [37]". In SP, however, we can explain the result quite simply and meaningfully. For simplicity, we confine the discussion to $\mathbb{R}^2$ and consider beamsplitters in just one- or two-particle interferometers. With each beamsplitter we associate a rotation $\Gamma$ of the unit circle in $\mathbb{R}^2$ (an unknown parameter associated with the beamsplitter).

(a) **Single-particle interferometer**: In a typical setup of this kind (as in Fig. 3 below in Sec. III) we have two 50-50 beamsplitters. The tangible particle travels first from a point-source to a beamsplitter [38] (as in Fig. 3), and from there it has a choice of paths to a second beamsplitter (in SP a shadow counterpart taking the alternative path). For $i = 1, 2$, we associate a rotation $\Gamma_i$ with the $i^{\text{th}}$ beamsplitter. If the exponentiated action of the tangible is of the form $r_1 e^{i\theta_1}$ when it encounters the first beam splitter, then it is transmitted if $\theta_1$ is in $\Gamma_1(\{(\cos\varphi, \sin\varphi): 0 \leq \varphi \leq \pi\})$, otherwise it is reflected. When the stream reaches the second beamsplitter, let $r_2 e^{i\theta_2}$ be the sum over the paths leading to the second beamsplitter, one path traveled by the tangible, the other, say, by its shadow counterpart. If $\theta_2$ is in $\Gamma_2(\{(\cos\varphi, \sin\varphi): 0 \leq \varphi \leq \pi\})$ the tangible is transmitted, otherwise it is reflected.

(b) **Two-particle interferometer**: There is a necessary positional uncertainty in the source for a two-particle interferometer [39]. Thus we have infinitely many paths, and the underlying space is non-measurable (as in Fig. 1 in Sec. I). This leads to a product of amplitudes on each side (Postulate 3), each separate amplitude being a sum over all possible paths in the related shadow stream. As in the single-particle case, the amplitude on a side (sum over two streams, as in Fig. 5 of Sec. V) is of the form $re^{i\theta}$. If the angle $\theta$ is in the interval $\Gamma(\{(\cos\varphi, \sin\varphi): 0 \leq \varphi \leq \pi\})$, the tangible is transmitted, otherwise it is reflected.

**Postulate 5 (Twin pairs)**: The emission of two entangled tangible particles traveling in opposite directions generates two oppositely directed shadow streams (oppositely directed by conservation of momentum). In a two-particle setup, these streams typically contain uncountably many particles accompanying the two twin daughter-tangibles. Moreover, we postulate that for each particle (tangible or shadow) in one of the paired streams, there is a twin counterpart in the other stream whose associated unit vector at the source event is pointing in the same direction as that of its twin. These are the identical, initial "instructions" carried by the twin particles, different pairs possibly pointing in different directions at the start. As the twin particles travel to oppositely directed destination points, they follow "action instructions" (described above).

In this paper we won't actually need the following postulate about spin, but for completeness we state it here (in terms of spin-½).

**Postulate 6 (spin-½)** Consider a spin-½ tangible particle, such as an electron, moving with its shadow counterparts through the inhomogeneous magnetic field of a Stern-Gerlach (SG) device. In SP we view each particle in the tangible's stream in terms of local realism (as in Fig. 3), picturing each particle as a tiny, spinning, top-like, dipole magnet. The tangible's shadow stream consists of infinitely many such "dipole tops" spinning in various arbitrary directions, a continuum of possibilities in $\mathbb{R}^3 \times SO(3)$, each top having corresponding simultaneous classical spin components in various directions.

For example, in Fig. 3 below we depict the tangible as the "darker" particle traveling in its shadow stream of infinitely many, paler, counterpart ideal particles.

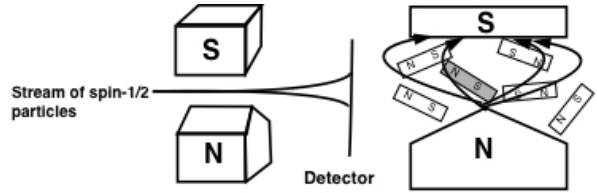

Fig. 3 The darker particle is a tangible particle traveling in a shadow homotopy class with other spin 1/2 particles (tiny dipole magnets)

The observables are the outcome of a sum of interference effects over all the paths associated with particles in the shadow stream, the tangible spinning in one of its two possible homotopy classes in the stream, where a spin path is in $\mathbb{R}^3 \times SO(3)$ (recall that $SO(3)$, whose fundamental group is of order 2, has just two homotopy classes of paths). The effects are quantified by the sum of exponentiated-action terms over the paths, where the action is a sum at each instant along a rotational path associated with a classical "top" action and a magnetic-field action [40].

There are just a small number of fundamental experimental results that must be accounted for by any mathematical representation of spin-½ [41, 42]. In standard quantum mechanics (SQM) this is done elegantly by a two-dimensional vector space over the complex numbers.

It is also done elegantly in SP by the path integral, in which we view spin in terms of a shadow stream in $\mathbb{R}^3 \times SO(3)$. When a spin-½ tangible particle enters an SG field for the first time, it will travel randomly in one of $SO(3)$'s two homotopy classes (for spin other than spin-½, one takes further projections onto the shadow stream). The sum of the exponentiated-action terms of all the particles in the tangible's homotopy class leads to one of the positions $+\hbar/2$ or $-\hbar/2$ on the SG detection plate. If, after traveling in a particular homotopy class, the tangible particle exits a prior SG field and then enters a field of the same orientation we postulate that it will continue in that homotopy class to produce the same outcome as previously. But if it enters a field oriented in a different direction it will go randomly into one of two homotopy classes for the new orientation, producing the usual result. If the tangible exits an SG device and then (while carrying a particular orientation) it enters a "modified-SG $\alpha$ device" (as defined by Feynman [43]), it will retain its previous orientation on exiting the magnetic field, because (in our explanation) the sum of "exponentiated-$\alpha$" terms in the two homotopy classes of the stream of the modified setup will recombine to the entering value, leaving the particle oriented in its previous direction. In this way we see that the SP outcomes replicate the fundamental quantum mechanical predictions. This is similar to how we accounted for the effects at a beamsplitter above.

### III. SINGLE (TANGIBLE)-PARTICLE INTERFERENCE

Feynman famously wrote about single-particle interference [44] (that is, single "tangible" in our framework): "… it contains the *only* mystery. We cannot make the mystery go away by 'explaining' how it works. We will just *tell* you how it works. In telling you how it works we will have told you about the basic peculiarities of all quantum mechanics." The SP approach, however, explains the mystery of single-particle interference in terms of counterpart ideal or shadow particles. This is similar to what Deutsch did in Ref. 1, but he viewed the process in terms of multiple worlds rather than in terms of paths connecting the initial and final (observed) event. In Sec. V we will also explain the mystery of the correlations for pairs of entangled particles.

Consider, for example, the following experimental setup shown in Fig. 4. Here a tangible particle travels in, say, a square-shaped interferometer, taking a fixed path from a point source to the first

11
beamsplitter, $BS_1$, from which it next has a choice of paths—"upper" or "lower".

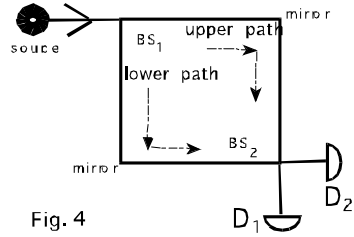

Fig. 4

At $BS_1$, 50% of the time the tangible will be transmitted and 50% of the time reflected (similarly for $BS_2$). When the paths along the sides form a square, as in the setup of Fig. 4, it is well-known that—as the particle travels to the second beamsplitter and on through to a detector—100% of the time it will be detected in $D_2$, never in $D_1$. This outcome is predicted in the standard two-dimensional vector algebra of SQM, a calculation that employs a basis of the eigenstates for a beamsplitter (associated with the operators R and T—"reflect" and "transmit"), etc. [Ref. 35]

We can also calculate the probability easily in SP by summing the exponentiated-action over the two paths to $BS_2$. Thus, associated with each stream going to a detector, we have (ignoring reflection) a final vector ($\approx e^{i\theta}$) a result of the fact that the action is the same over congruent paths. Then we add the probabilities over each possible path to a detector, but now we insert a factor of $e^{i\pi/2}$ for each instance of reflection in the final sum:

$$\text{Amplitude for arrival at } D_1 \text{ is} \approx \underbrace{e^{i\theta}e^{i\pi/2}}_{\text{top}} + \underbrace{e^{i\theta}e^{i\pi/2}e^{i\pi/2}e^{i\pi/2}}_{\text{bottom}} = e^{i\theta}e^{i\pi/2}(1+e^{i\pi})$$

$= e^{i\theta}e^{i\pi/2}(1+-1) = 0$. So the absolute square (the probability) is 0.

And

$$\text{Amplitude for arrival at } D_2 \text{ is} \approx \underbrace{e^{i\theta}e^{i\pi/2}e^{i\pi/2}}_{\text{top}} + \underbrace{e^{i\theta}e^{i\pi/2}e^{i\pi/2}}_{\text{bottom}} = 2e^{i\theta}e^{i\pi}.$$

So the absolute square is proportional to $e^{i\theta}e^{i\pi}e^{-i\theta}e^{-i\pi} = 1$.

Although SP and SQM yield the same probability for quantum events, there is a crucial difference: In SQM, as noted earlier, any two wavefunctions that differ by a complex scalar or phase factor represent the same physical reality. On the other hand, in SP by Postulate 4a, when the tangible arrives at the second beamsplitter, the vector sum of the shadow streams over "upper" and "lower" paths in Fig. 4 above is taken to be pointing in a fixed direction (that is, it is associated with a unique phase factor). This is the difference between SQM and SP for single-particle interference. The assumed fixed (but unknown) direction of the sum-vector in SP for the two streams (where each stream might contain only a single particle in the case of Fig. 4) does not affect the probability calculation, because the absolute square is 0 for arrival at $D_1$ and 1 for $D_2$. The assumption of counterpart ideal particles is all that is needed to explain the outcome for single-particle interference. But the direction of the sum-vector does play an important role for a pair of entangled particles in Sec. V.

There are various paradoxes involving single-particle interference. We will examine two, showing how easily they are eliminated in SP. Consider, first, John Wheeler's celebrated "delayed-choice" paradox. In this gedanken experiment one imagines a tangible particle being sent from a source through a beamsplitter in a galactic-sized interferometer of the type in Fig. 4. (Wheeler chose the astronomical scale to dramatize the resulting paradox). Here, a tangible photon enters the apparatus at $S$ and is sent by the beamsplitter along either the upper or lower path to the second beamsplitter and thence to one of the detectors, $D_2$ or $D_1$.

As we saw above, when the path lengths are suitably arranged—as in Fig. 4 where the sides of the interferometer are congruent—constructive interference will cause the particle to be detected always at one location, $D_2$, whereas destructive interference will prevent its detection at $D_1$. But if we peek at the photon while it is in transit to locate in which arm of the interferometer the particle is traveling, even long after it has set out on its journey from, say, a distant star, we will destroy the interference, allowing the particle to reach the second detector $D_1$ some of the time.

In discussing Wheeler's paradox the distinguished mathematical physicist Roger Penrose writes [45],

> "… the key puzzle is that somehow a photon (or other quantum particle) seems to have to 'know' what kind of experiment is going to be performed upon it well in advance of the actual performing of that experiment. How can it have the foresight to know whether to put itself into 'particle mode' or 'wave mode' as it leaves the (first) beamsplitter."

But, as noted earlier, it's easy to explain the paradox in SP using ideal elements: When we look at the photon before it reaches the second beamsplitter, we disrupt one of the two shadow streams traveling from the first to the second beamsplitter, which alters the interference effects leading to what we see at the detectors.

Another famous single-particle paradox is what's known as "interaction-free measurement" (IFM), said to be "A novel manifestation of non-locality of quantum mechanics … [showing] that it is possible to ascertain the existence of an object in a given region of space without interacting with it. [46]" Thus suppose that an ultra-sensitive bomb (one that can be triggered by a single photon) is placed in the lower arm of the interferometer. If the bomb is a dud, then a photon (tangible or shadow in SP) will pass through the bomb unscathed and produce interference. Thus, by the design of the interferometer if the bomb is a dud then only the $D_2$ detector will fire (as we calculated above).

On the other hand, suppose the bomb is *not* a dud. Then if the tangible photon takes the lower path it will trigger the bomb, which would occur roughly half the time: destroying 50 percent of the usable bombs. But for the other 50 percent of the time, on average, the first beamsplitter will send the photon along the upper path. In such cases there will be no interference at the second beamsplitter because—as explained by SP—the shadow particle moving along the lower path is blocked. Therefore, in those cases in which the tangible photon takes the upper route, half the time the $D_1$ detector will fire; and in such cases (25 percent of the time overall) we know the bomb is not a dud, without having destroyed it—a desirable result for bomb manufacturers. But also one that is likely (as Kwait et. al point out [47]) to have important technical ramifications in photography, medicine, and various other areas of research.

The IFM result has been called "quantum seeing in the dark [48]." Clearly, it is a remarkable property of quantum mechanics. From our viewpoint, however, it might be called more aptly, "seeing with shadow particles." In SP it is certainly no longer a paradox.

Our Postulate 4 in Sec. II (regarding the direction of the sum-vector) is clearly compatible with experiment, since experiment and SQM are silent on this matter, as noted in the Feynman quote of Sec. II. Of course the shadow stream of Deutsch's counterpart particles was never employed by Feynman, who was working within the framework of SQM, whereas Deutsch was working in MWI. Indeed, there can be no local realism (as in Postulate 4) in the SQM framework, as Bell's argument in fact shows (Feynman was well aware of Bell's theorem [49]). But, although Feynman never described his construct in terms of a shadow stream, his writings are studded with statements compatible with our approach, such as the particle "smells all the paths in the neighborhood [50]." (For "smells all paths" to hold for a single particle, the particle would have to travel paradoxically at infinite speed, since there are uncountably many paths in a finite region of space; in SP we avoid this paradox by assuming a one-to-one correspondence between particles and possible paths in the stream.) Also "… light comes in this form—particles. It is very important to know that light behaves like particles, especially for those of you who have gone to school, where you were probably told something about light behaving like waves. I'm telling you the way it does behave—like particles."[51]

## IV. A LOCALITY CONDITION BASED ON EXPONENTIATED ACTION

*As is clear from Sec. I, our argument that SP is a local system cannot apply the Bell factorizability condition of Eq. (1).* As pointed out in Sec. I, this condition is meaningless in the space of all paths, as are the subsequent steps in Bell's argument that use translation invariance. Thus in our system the factorizability condition involving a probability distribution is not well-defined in SP and would involve the conflation of a logical operation over separate mathematical structures or domains, one domain being measurable and the other nonmeasurable. There is, however, another simple mathematical condition for locality that we can use in our approach—one that is every bit as representative of classical locality as the factorizability condition. This condition is based on the fact that in SP we can associate a unit vector or "clock-pointer," $\exp(iS[path]/\hbar)$, with each particle in a shadow stream as the particle travels its path from the source to a detector (as noted in Sec. II).

To help explain how this condition applies to our framework, we will now describe a gedanken, toy system in which two classical objects (billiard balls, tennis balls, bullets, etc.) proceed out from a source, traveling in opposite directions to forks located at equal distances along each path, the fork on each side leading to upper and lower branches. This is a local system, with just two paths from the source to the branch points on each side (unlike in the quantum mechanical case, where there are infinitely many such paths).

In this setup we imagine that each object carries a "clock-pointer," $\exp(iS[path])$, as the object on each side travels its path out from the source. The two pointers are set randomly at the source event, but point initially in the same direction. We also suppose that the pointers rotate counterclockwise at identical rates (given by the classical action—notice that we do not need the factor of $1/\hbar$ in this classical example) as the objects travel along their respective paths (in this classical setup there just one path on each side). Just before the objects reach a fork point, "measurements" are carried out—$\alpha$ on the left side, $\beta$ on the right, where $\alpha$ and $\beta$ are one of the angles $0, 2\pi/3, 4\pi/3$. For example, a setting of $\alpha$ ($0 < \alpha < 2\pi$) slightly lengthens the path of the particle on the left so that its clock-pointer further rotates by the additional amount $\alpha$ (the angle $\alpha$) over that stretch of the path; and similarly for $\beta$ on the other side. When the objects reach a fork, identical bivalent functions send them along the upper or lower branches at the fork, depending on whether their clock-pointers are pointing between $0$ and $\pi$, or between $\pi$ and $2\pi$.

Because the action is the same over congruent paths (a point that holds also in the quantum mechanical case below), the correlation of position between the sides—the objects ending up on corresponding forks: (upper, upper) or (lower, lower)—is 100% when $\alpha = \beta$. When $\alpha \neq \beta$, the probability of correlation is 1/3, as shown in Appendix A. Here we note that for, say, $\alpha > \beta$ when the object arrives at a fork, the pointer on the right side is directed at some angle $\gamma$, while on the left side it is at angle $\gamma + (\alpha - \beta)$. The toy system obviously operates in a measurable space. Thus the correlations, being locally explicable, must satisfy Bell's inequality, but the toy system illustrates why the outcome on, say, the left side in a local theory can involve the measurement $\beta$ on the right side (which might at first glance seem contrary to locality). This is because the outcome is accomplished by congruent paths (apart from the measurements) and by clock-pointers—built-in "genetic" instructions. As we show below, this condition transfers also to the quantum setup, which has infinitely many possible paths. Because of the clock-pointers, there is no need for information to be sent across the origin in order to coordinate the correlations. (Interestingly, the value 1/3 may be a lower bound for a classical system incorporating the above measurements: It is considerably less, for instance, than the value of 1/2 for Mermin's famous red-green version of Bell's argument [52], which employs similar measurement settings.)

## V. A COUNTEREXAMPLE TO BELL'S THEOREM BASED ON THE HIDDEN REALITY OF SHADOW PARTICLES

In our analysis of single-particle interference in Sec. III, locality and non-measurability were not concerns. This is no longer the case for two-particle interference, where Bell's theorem is widely believed to rule out local realism.

Recall that in Fig. 1 of Sec. I we compared the conventional picture of a Rarity-Tapster two-particle interferometer, on the left, to that of the SP version on the right. In a Rarity-Tapster interferometer, two entangled particles are generated by down conversion at the source. They then travel to mirrors on each side, from which they are reflected to 50-50 beamsplitters at $A$ and $A'$, and then on to detectors $u$, $d$ on the left and $u'$, $d'$ on the right. On each side there are phase-shifters—$\alpha$ on the left, $\beta$ on the right.

As pointed out in Sec. I, in any two-particle interferometer, interference can only occur if there is a certain amount of positional uncertainty in the source, unlike with single-particle interference, which requires a point source [53]. Recall that the Rarity-Tapster device handles this as follows: At each run of the experiment, paired spinless particles in SP travel first to floor- and ceiling-mirrors (M-M′ lower and M-M′ upper in Fig. 5 below), from which they are then, on the respective sides, reflected to 50-50 beamsplitters at $A$ and $A'$. The setup generates infinitely many possible paths, forming a non-measurable space in SP. In SP this results in four infinite streams of particles X, Y, X′, and Y′.

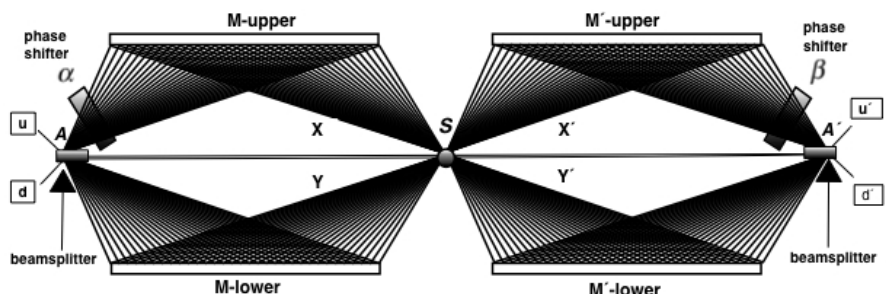

Fig. 5  SP picture showing the hidden reality underlying the conventional picture of a Rarity-Tapster interferometer

15As in Sec. III, we of course know in advance that the two systems, SQM and SP, yield equivalent outcomes. But in SQM the mathematical implementation of the Rarity-Tapster interferometer is based on a finite-dimensional Hilbert space of square-integrable wavefunctions—a measurable space (unlike in SP). In fact, the conventional picture of the interferometer (on the left side of Fig. 1 in Sec. I) seems to suggest just finitely many paths (though of course there are infinitely many kets or wavefunctions in the vector space associated with the SQM representation). On the other hand, there are infinitely many possible paths indicated in the SP picture in Fig. 5 above.

Without loss of generality we can confine our discussion to a planar setup in $\mathbb{R}^2$ (a consequence of the mathematical properties of the action functional [54]).

As noted in Sec. I, in the mathematical representation based on the SQM vector-space approach, one usually begins by defining the quantum state of the pair of entangled particles as, say,

$$|\psi\rangle = (|x\rangle_L |y'\rangle_R + |y\rangle_L |x'\rangle_R)/\sqrt{2},$$

where (as in the left-side of Fig. 1 in Sec. I) the ket $|x\rangle_L$ denotes the particle on the left traveling along path $x$, etc.

But in SP it is convenient from the outset to work with amplitudes for joint detection. Hence we let $\lambda = L \cup R$ represent the set of infinitely many possible paths going from the source to the two separated systems represented by $A$ and $A'$ at the beamsplitters in Fig. 5, with $L = X \cup Y$ and $R = X' \cup Y'$. Thus (as noted in Sec. I) $\lambda$ is the set of "hidden variables" comprising the possible paths $x(t)$ from the source to the beamsplitters. As we will show, $\lambda$ constitutes the hidden reality underlying the quantum mechanical observables, accounting for the correlations in terms of local realism.

An expression such as $\langle A | X \rangle_L$ below denotes the amplitude for a stream of particles traveling along paths in the upper left stream from the source $S$ to the beamsplitter at $A$ (here the measurement setting $\alpha$ is left implicit). The amplitude $\langle A | X \rangle_L$ is proportional to the sum

$$\sum_{\substack{\text{over all upper paths} \\ x(t) \text{ from } S \text{ to } A}} exp(iS[x(t)]/\hbar)$$

Similarly for the other amplitudes. We will now illustrate in SP one possible approach to computing the probability for the correlations between the sides—the outcome being equivalent to that of SQM.

Now in any two-particle interferometer similar to ours, the emission angle subtended at the source by the upper paths must be sufficiently acute for correlations to occur [55]. That is, if the angle is too large, then each side corresponds to single-particle interference, and the correlations disappear. Also, in SP a tangible particle always pursues a single (randomly taken) trajectory by Postulate 1. This means that there are just two choices for what may be termed the "upper" tangible particle: It can only take a path in the homotopy class $X$ or in the class $X'$ in the SP path-setup of Fig. 5— which path it does take varies randomly by Postulate 1 at each trial of the experiment. If the upper tangible takes, say, a path in $X$ then the other, lower tangible must by conservation of momentum take a linked, oppositely directed path in $Y'$. Similarly, if the upper takes a path in $X'$, then the other tangible is linked to take a path in $Y$. This implies that the particles taking paths in the homotopy classes $X$ and $X'$ are from different streams; likewise for those taking paths in $Y$ and $Y'$. In other words, $X$ and $Y$ are generated by different tangibles (in our set of two entangled tangible particles); likewise so are $X'$ and $Y'$. By Postulate 3, the composite amplitude of the shadow streams generated by different tangibles is the product of the separate amplitudes. Therefore, factoring in $i$ (= $\exp(i \cdot \pi/2)$) for reflection from a beamsplitter, we have,

$$\text{both "up" detectors flash:} \quad i \cdot \underbrace{\langle A | X \rangle_L}_{\substack{\text{left top path to up} \\ \text{detector}}} \cdot \underbrace{\langle B | Y' \rangle_R}_{\substack{\text{right bottom path} \\ \text{to up detector}}} + \underbrace{\langle B | X' \rangle_R}_{\substack{\text{right top path to} \\ \text{up detector}}} \cdot i \cdot \underbrace{\langle A | Y \rangle_L}_{\substack{\text{left bottom path} \\ \text{to up detector}}}, \quad (2)$$

$$\text{both "down" detectors flash:} \quad \underbrace{\langle A | X \rangle_L}_{\substack{\text{left top path to} \\ \text{down detector}}} \cdot \underbrace{\langle B | Y' \rangle_R}_{\substack{\text{right bottom path} \\ \text{to down detector}}} \cdot i + \underbrace{\langle B | X' \rangle_R}_{\substack{\text{right top path to} \\ \text{down detector}}} \cdot \underbrace{\langle A | Y \rangle_L}_{\substack{\text{left bottom path to} \\ \text{down detector}}} \cdot i. \quad (3)$$

Using the congruence of the paths (apart from the phase shifts, $\alpha$ and $\beta$) we can represent these amplitudes (which are just complex numbers) by

$$\langle A | X \rangle_L = re^{i\theta}e^{i\alpha}, \langle A | Y \rangle_L = re^{i\theta}, \langle B | Y' \rangle_R = re^{i\theta}, \langle B | X' \rangle_R = re^{i\theta}e^{i\beta}. \quad (4)$$

Thus the amplitude that both detectors flash the same is

$$ire^{i\theta}e^{i\alpha} \cdot re^{i\theta} + re^{i\theta}e^{i\beta}i \cdot re^{i\theta} = ir^2 e^{i2\theta}(e^{i\alpha} + e^{i\beta}). \quad (5)$$

When we take the absolute square (multiply by the complex conjugate), we have

$$P(\text{both-flash-same over } \lambda | \alpha, \beta) \propto r^4(e^{i\alpha} + e^{i\beta})(e^{-i\alpha} + e^{-i\beta}) = r^4(2 + e^{-i(\alpha-\beta)} + e^{i(\alpha-\beta)}). \quad (6)$$

Performing simple algebraic operations over the complex numbers, and choosing a suitable proportionality factor, we see that the probability for the detectors to signal the same, both "up" or both "down," is thus (as in SQM),

$$\cos^2 \frac{(\alpha - \beta)}{2}. \quad (7)$$

Now expressions such (2) or (3) seem to suggest that the tangible particles must somehow communicate across the source in order to coordinate the predicted correlations. But of course by itself such an expression can show no such thing. A conclusion of that kind requires Bell's argument, which breaks down in SP. We will now show that the correlations occur, in fact, by local interaction on each side of experimental device.

Because SP is a nonmeasurable space, our argument that SP is a local system cannot be based on the Bell factorizability condition, but (as noted in Sec. IV) there is another simple mathematical approach based on "clock-pointers," exponentiated-action terms. The following approach parallels the one for the toy system in Sec. IV involving classical objects.

Note that each path in the stream in the lower half on the left side Fig. 5 corresponds in a one-to-one fashion to a congruent path in the lower half of the right side. Congruent paths have equivalent exponentiated actions, so the sum on the left side over all the lower paths will be the same as on the right. That is, $\langle A|Y \rangle_L = \langle B|Y' \rangle_R$. Therefore (substituting "equals for equals" into expression (2) or (3)), the amplitude for both detectors to signal the same (that is, both "up" or both "down") is,

$$i\left(\langle A|X \rangle_L \langle B|Y' \rangle_R + \langle B|X' \rangle_R \langle A|Y \rangle_L\right) = i\left(\langle A|X \rangle_L \langle A|Y \rangle_L + \langle B|X' \rangle_R \langle B|Y' \rangle_R\right). \tag{8}$$

A product such as $\langle A|X \rangle_L \langle A|Y \rangle_L$ on the right half of Eq. (8) is of the form $re^{i\theta}$, for some $\theta$. By postulate 4 in SP the local interference effects associated with such an amplitude at the beamsplitter are what determine where the stream containing the tangible particle goes next after leaving the beamsplitter—that is, "up" or "down" in our setup. This is an independent physical process on each side. (Indeed it closely resembles the similar situation in the toy example of Sec. IV, which obviously displays local realism.)

In the two-particle interferometer of Fig. 5, the interference on a side comes from a product of two local amplitudes—such as $\langle A|X \rangle_L \langle A|Y \rangle_L$ on the left side of the SP picture in Fig. 5—a composite amplitude, but of the form $re^{i\theta}$. This composite amplitude is independent of the similar one $\langle B|X' \rangle_R \langle B|Y' \rangle_R$ on the other side by postulate 4. The sum

$$\langle A|X \rangle_L \langle A|Y \rangle_L + \langle B|X' \rangle_R \langle B|Y' \rangle_R \tag{9}$$

is therefore the analogue in SP of the Bell factorizability condition (Eq (1) in Sec. I).

Thus, although Bell's argument breaks down in SP, we see that an intuitively appealing "local condition" (expression (9)) nevertheless holds in the space of all paths, while at the same time the predicted correlations violate Bell's famous inequality. The dynamics are similar to what takes place in the toy, classical system of Sec. IV, in which we imagined that a pair of "twin" classical objects travel in opposite directions from a source along a single path on each side. The tighter quantum mechanical results (that is, a smaller probability) stem, however, from the fact that the sum (and then product) over exponentiated-action terms issuing from interference effects leads to a complex number such as $re^{i\theta}$ (whereas there are no interference effects in the classical, single-path system).

To recapitulate: When entangled particles travel out from a source event along a path, their "clock-pointers," $\exp(iS[path]/\hbar)$, are set pointing in the same direction (postulate 5). This is the initial instruction, the same on both sides. As the upper and lower streams of particles (tangible and shadow) on each side travel to a beamsplitter, their clockpointers over congruent paths (governed by identical action-instructions) rotate at the same rate. In this way, the particles in a stream follow blindly their initial programming—their "genetic action-instructions," together with the path-dynamics governed by the laws of physics pertaining to the action over a path. By Postulate 4 the effects at the beamsplitters are governed by the direction of the sum-vector resulting from a local sum of paths in the product of two amplitudes— $\langle A|X \rangle_L \langle A|Y \rangle_L$ on the left side, and $\langle B|X' \rangle_R \langle B|Y' \rangle_R$ on the right. This is the SP analogue of the Bell factorizability condition for locality (Eq. (1) in Sec. I). In other words, the outcome on each side is the result of local interaction involving exponentiated-action terms over all paths on a side, and is separate from what happens on the other side. Thus no information needs to be sent across the source. The correlated outcomes between the sides are the consequence of various correspondingly congruent paths, leading to correspondingly equivalent exponentiated-action terms, apart from the phase-shift measurements $\alpha$ and $\beta$. Each phase shift—$\alpha$ on the left, and $\beta$ on the right—affects just its side independently, as in the toy system of Sec. IV, leading to the joint probability of the correlations. **QED**

The same construction as the one above works also for spin, although the details are outside the scope of the present paper. Here is a quick sketch of the idea for spin ½. As pointed out earlier (Sec. II), the paths for spin are "rotational" paths in $\mathbb{R}^3 \times SO(3)$, and the action over a path is the sum of a classical top-action and a magnetic-field action [56]. Thus, when entangled spin ½ particles enter Stern-Gerlach (SG) devices on each side of the experiment, the correspondingly congruent spin paths, apart from the measurements, coordinate the correlations, as with the congruent paths in the two-particle interferometer (the system described in Postulate 6 of Sec. II). Here, $SO(3)$ has two homotopy classes of paths (its fundamental group is of order 2); and, as is well known, we can model $SO(3)$ homeomorphically in terms of a solid ball of radius $\pi$, with antipodal points identified, where a point a distance $\gamma$ from the center of the ball and along a radius pointing in the direction of the unit vector $\boldsymbol{n}$ corresponds to a counterclockwise rotation $\gamma$ about the axis $\boldsymbol{n}$. One can thus project over paths along, say, the $y$-axis within the ball (where we are assuming that the entangled particles are traveling in opposite directions from a source, with rotation measurements $\alpha$ and $\beta$ performed around the $y$-axis). In this fashion one can carry out an analysis similar to the above interferometer argument.

We also note that Bell type arguments (such as GHZ [57]) for more than two entangled particles cannot be carried out in SP. This is important because, unlike Bell's theorem, the more-than-two-particle arguments are not statistical in nature. Although a detailed justification is, as for the above spin-½ argument, outside the scope of this paper, here is a quick sketch of why GHZ fails in the space of all paths. It hinges on the notion of an "observable," which in SQM is the eigenvalue of a self-adjoint operator.

Now SQM and SP have the same observables, but in SP the element of reality of the tangible particle is not the SQM observable, which is the result at measurement of interference with infinitely many other particles in the stream. The SP observable prior to measurement is the spin of the tangible particle. That is, the element of reality in SP is the actual top-like spin of the tangible particle (illustrated in terms of local realism by the darker colored object in Fig. 3 of Sec. II) along the tangible's actual path as the particle exits the SG field. But at measurement what we observe (the SQM observable) is the result of the contribution from an entire manifold of paths neighboring that of the tangible particle. The observable merely "points to" the element of reality, as it resides in a swirl of interference effects. The tangible's exact path is what is associated at each instant with the tangible's classical property, its "SP element of reality" (as in Fig. 3 of Sec. II).

On the other hand, the GHZ argument identifies EPR's element of reality with the notion of an SQM observable. This identification easily leads to a contradiction, as GHZ shows. Yet, EPR never said an element of reality was identical to an observable (they gave a sufficient condition based on 100 percent predictability—which is consistent with how we just described the concept as a pointer to the hidden reality of the tangible particle in SP—its actual classical spin prior to measurement). But whatever ESP actually meant by an element of reality, the GHZ argument fails in SP because, among other things, path configurations do not form a number field in which one can carry out arithmetic operations such as multiplication, etc.—which is how GHZ was able to predict the outcome of various so-called "intrinsic spin" observables given the results of others.

Incidentally, in certain respects the notion of an observable in SP resembles Bell's idea of a "beable," where he writes: "The beables of the theory are those elements which might correspond to elements of reality, to things which exist. Their existence does not depend on 'observation'."[58]

### VI. CONCLUSION

The consequences of the breakdown of the factorizability condition for locality are, so we believe, profound for quantum foundations. In the nearly fifty years since Bell's original argument, many thousands of articles, in both the technical and popular literature, have been published in which Bell's condition of locality has played a prominent role. On the other hand, our implementation of Feynman's formulation of quantum mechanics employing Deutsch's ideal particles is easily seen to replicate the predictions of quantum mechanics in a locally realistic fashion.

The space of all paths is a simple system that duplicates the predictions of standard quantum mechanics, while eliminating paradoxes. Moreover, unlike SQM, it is not opaque as to the inner mechanism underlying the quantum observables.

There is also this: Essentially there are just two, mutually exclusive, choices for quantum foundations: systems structurally similar to the space of all paths or systems that harbor action at a distance. One can see this by building on an important result contained in a paper of Bernstein, Greenberger, Horne, and Zeilinger (BGHZ), "Bell theorem without inequalities for two spinless particles" [59]. For their proof to go through, BGHZ found it necessary to augment it with a premise called "emptiness of paths not taken," EPNT. Of course SP is, in a sense, the denial of EPNT. By extending the EPNT hypothesis to include homotopy classes of paths, one has, mutatis mutandis, BGHZ's "action at a distance" (to use a phrase from Einstein that BGHZ did not employ). Although one could deny EPNT and still have a nonlocal theory (for example, David Bohm's pilot-wave theory, which in effect "fills" the paths with a "pilot wave") [60], we nevertheless have the result that "SP" implies "no action at a distance," and "not SP" implies "action at a distance."

Unfortunately, SQM, despite its great predictive power, has been riddled with paradoxes since its inception. This has caused many physicists to retreat from the natural view of science as offering an explanation of phenomena. A quite striking instance of this involves entangled particles, where the predicted correlations are taken to be completely inexplicable [61], though one might claim that one of the fundamental purposes of a scientific theory is to explain correlations, as science does for example in biology with the theory of evolution.

Indeed, as Bell noted [62], "the scientific attitude is that correlations cry out for explanation." As Bell told Jeremy Bernstein [Ref. 62], the predicted correlations for two entangled particles seem to demand something like the "genetic" hypothesis—where identical twins carry with them identical genes. "This is so rational that I think that when Einstein saw that, and the others refused to see it, *he* was the rational man … The other people, although history has justified them, were burying their heads in the sand … So for me, it is a pity that Einstein's idea doesn't work. The reasonable thing just doesn't work."

On the contrary, as we have just seen, the reasonable thing does in fact work—in SP. As we saw in Sec. III, SP eliminates easily single-particle interference paradoxes in an intuitively appealing framework similar to that of MWI, but one which seems considerably simpler than MWI with its universal wavefunction, etc. Furthermore, SP (unlike MWI) also eliminates multi-particle paradoxes (as we saw in Sec. V) via a form of genetic hypothesis—identical action instructions.

We end with a quote from Einstein [Ref. 62]:

> I hope that someone will discover a more realistic way, or rather a more tangible basis than it has been my lot to find. Even the great initial success of the quantum theory does not make me believe in the fundamental dice-game, although I am well aware that our younger colleagues interpret this as a consequence of senility. No doubt the day will come when we will see whose instinctive attitude was the correct one.

## APPENDIX A

In computing the probability for the correlations in Sec. IV when $\alpha \neq \beta$, we suppose, say, $\alpha > \beta$, where of course by design each measurement is less than $2\pi$. Then, after the measurements, the objects arrive at the fork points with the clock-pointer on the right at some angle $\gamma$ while on the left it's at $\gamma + (\alpha - \beta)$, where $\alpha - \beta = 2\pi/3$ or $4\pi/3$. The probability that the objects end up on corresponding branches is (because of the bivalent functions $A$ and $B$) the probability that both pointers are between 0 and $\pi$, or both are between $\pi$ and $2\pi$. This probability in both cases is 0 when $\alpha - \beta = 4\pi/3$. When $\alpha - \beta = 2\pi/3$ we have, as noted, two possibilities: that both pointers are between 0 and $\pi$ or both are between $\pi$ and $2\pi$. Observe first that

$$\Pr(0 < \gamma < \pi/3) = (\pi/3)/2\pi = 1/6.$$

The product that both pointers are between 0 and $\pi$ is the product:

$$\Pr(0 < \gamma + 2\pi/3 < \pi) \cdot \Pr(0 < \gamma < \pi) = \Pr(0 < \gamma < \pi/3) \cdot \Pr(0 < \gamma < \pi) = (1/6)(1/2).$$

The same holds for the case where $\pi < \gamma < 2\pi$, and thus the probability is 1/6 when $\alpha > \beta$. Similarly, when $\beta > \alpha$. Therefore when $\alpha \neq \beta$ the total probability is 1/3.

---

[1] D. Deutsch, *The Fabric of Reality*, Penguin Press, 1998

[2] R.P. Feynman, "Space-Time Approach to Non-Relativistic Quantum Mechanics," Rev. Mod. Phys. 20, 367-387 (1948) reprinted in *Feynman's Thesis, A New Approach to Quantum Mechanics*, edited by Laurie M. Brown, World Scientific Publishing, 2005.

[3] L. Vaidman, "Many Worlds Interpretation of Quantum Mechanics," Stanford Encyclopedia of Philosophy, 2002

[4] T. Maudlin, Am. J. Phys. **78** (3), January 2010

[5] J. S. Bell, Physics I, 195-200 (1964) (reprinted in J. S. Bell, *Speakable and Unspeakable in Quantum Mechanics*, 2nd ed., Cambridge University Press, 2004 )

[6] A. Zeilinger, et al., Nature, Vol 446,19m April 2007| doi:10.1038/05677

[7] G. W. Johnson and M. L. Lapidus, *The Feynman Integral and Feynman's Operational Calculus*, Oxford Mathematical Monographs, Oxford University Press (2002), p. 32

[8] J. S. Bell, "Bertlmann's socks and the nature of reality," *Speakable and Unspeakable in Quantum Mechanics*, 2nd ed., Cambridge University Press, 2004

[9] Y. Zhang, S. Glancy, and E. Knill, "Asymptotically optimal data analysis for rejecting local realism," Phys. Rev. A (accepted Nov 21, 2011)

[10] A. Acín, S. Massar, S. Pironio, Phys. Rev. Lett. **108**, 100402 (2012) – published March 9, 2012

[11] J. S. Bell, "On the Problem of Hidden Variables in Quantum Mechanics", Reviews of Modern Physics vol. 38, number 3, july 1966

[12] W. C. Myrvold, Philosophy of Science 70 (December 2003)

[13] M. Horne, A. Shimony, A. Zeilinger, Nature Vol 347, 4 October 1990

[14]

[15] D.M. Greenberger, M. A. Horne, A. Shimony, A. Zeilinger, Am. J. Phys. 58 (12), December 1990

[16] D. M. Greenberger, M. A. Horne, A. Zeilinger, Physics Today, August 1993

[17] S. Weinberg, *Lectures on Quantum Mechanics*, Ch. 9, Cambridge University Press, 2012

[18] J. Keisler, *Elementary Calculus An Infinitesimal Approach*, Prindle, Weber and Schmidt, 1976

[19] A. Robinson, *Non-standard analysis*, North-Holland Publishing Co., Amsterdam 1966.

[20] M. Chaichian, A Demichev, *Path Integrals in Physics*, Vol 1, CRC Press, January 2001

[21] R. Feynman, A. Hibbs, *Quantum Mechanics and Path Integrals*, Ch 2, Dover Books, emended edition 2005

[22] L. Schulman, *A Path Integral for Spin*, Phys. Rev. 176, 1558-1569 (1968)

[23] G. Jaeger, *Quantum Information: An Overview*, p. 62

[24] V. Scarani, A. Suarez, Am. J. Phys. **66** (8), August 1998

[25] J. S. Bell, "Bertlmann's socks and the nature of reality," *Speakable and Unspeakable in Quantum Mechanics*, 2nd ed., Cambridge University Press, 2004

[26] M. Hoban, J. Wallman, and D. Browne, "Generalized Bell Inequality Experiments and Computation," Phys. Rev. A **84**, 062107 (2011), Dec. 6

[27] D.M. Greenberger, M. A. Horne, A. Shimony, A. Zeilinger, Am. J. Phys. 58 (12), December 1990

[28] D. Morin, *Introduction to Classical Mechanics: With Problems and Solutions*, Cambridge University Press, 2007, Ch. 6

[29] A. Atland, B. Simmons, *Condensed Matter Field Theory*, Cambridge University Press, 2006, Sec. 3.3.5

[30] D. Morin, *Introduction to Classical Mechanics: With Problems and Solutions*, Cambridge University Press, 2007, Ch. 6

[31] H. Kleinert, *Path Integrals in Quantum Mechanics, Statistics, Polymer Physics, and Financial Markets*, 2009, World Scientific 5th edition

[32] M. Gardner, J. A. Wheeler, *Quantum Theory and Quack Theory*, New York Review of Books, Vol. 26, Number 8, May 17, 1979

[33] R. P. Feynman, QED: *The Strange Theory of Light and Matter*, Princeton University Press, Princeton, NJ, 1986. An excellent elementary discussion of Feynman's approach is also in J. Ogborn and E. F. Taylor "Quantum physics explains Newton's laws of motion," Physics Education, January, 2005

[34] R. P. Feynman, R. B. Leighton, and M. Sands, *The Feynman Lectures on Physics* (Addison-Wesley, Reading, MA, 1964), Vol. III, Ch 3

[35] M. O. Scully, K. Druhl, Phys. Rev. A 25, 2208-2213 (1982)

[36] V. Scarani, A. Suarez, Am. J. Phys. **66** (8), August 1998

[37] R. P. Feynman, QED: *The Strange Theory of Light and Matter*, Princeton University Press, Princeton, NJ, 1986.

[38] V. Scarani, A. Suarez, Am. J. Phys. **66** (8), August 1998

[39] D. M. Greenberger, M. A. Horne, A. Zeilinger, Physics Today, August 1993

[40] A. Atland, B. Simmons, *Condensed Matter Field Theory*, Cambridge University Press, 2006, Sec. 3.3.5

[41] J. S. Townsend, *A Modern Approach to Quantum Mechanics*, Ch 1, University Science Books, 2000

[42] R. P. Feynman, R. B. Leighton, and M. Sands, *The Feynman Lectures on Physics* (Addison-Wesley, Reading, MA, 1964), Vol. III, Ch 5

[43] R. P. Feynman, R. B. Leighton, and M. Sands, *The Feynman Lectures on Physics* (Addison-Wesley, Reading, MA, 1964), Vol. III, Ch 5

[44] Feynman, R. P.; Leighton, R. B.; Sands, M. *The Feynman Lectures on Physics, Vol. 3*; Addison-Wesley: Reading, 1965, p. 1-1.

[45] R. Penrose, *The Road to Reality : A Complete Guide to the Laws of the Universe* Alfred A. Knopf, 2005

[46] A. C. Elitzur , L. Vaidman  *Quantum Mechanical Interaction-Free Measurements*, Found. Phys.,Vol. 23, No. 7,  987–997, July (1993)

[47] P.G. Kwiat, H. Weinfurter, A. Zeilinger, *Quantum Seeing in the Dark*, Scientific American **275**, 52-58 (1996).

[48] P.G. Kwiat, H. Weinfurter, A. Zeilinger, *Quantum Seeing in the Dark*, Scientific American, November, 1996

[49] Feynman letter to David Mermin, quoted by L. Gilder, *The Age of Entanglement: When Quantum Physics Was Reborn*, p. 20, Alfred A Knopf, 20008

[50] R. P. Feynman, R. B. Leighton, and M. Sands, *The Feynman Lectures on Physics* (Addison-Wesley, Reading, MA, 1964), Vol. II, Ch 19

[51] R. P. Feynman, QED: *The Strange Theory of Light and Matter*, Princeton University Press, Princeton, NJ, 1986.

[52] N. D. Mermin, *Bringing home the atomic world: Quantum mysteries for anybody*, Am. J. Phys. 49 (10),Oct. 1981

[53] D. M. Greenberger, M. A. Horne, A. Zeilinger, Physics Today, August 1993

[54]  D. Morin, *Introduction to Classical Mechanics: With Problems and Solutions, Cambridge University Press*, 2007, Ch. 6

[55] D. M. Greenberger, M. A. Horne, A. Zeilinger, Physics Today, August 1993

[56] A. Atland, B. Simmons, *Condensed Matter Field Theory*, Cambridge University Press, 2006, Sec. 3.3.5

[57] D.M. Greenberger, M. A. Horne, A. Shimony, A. Zeilinger, Am. J. Phys. 58 (12), December 1990

[58] J. S. Bell, *Speakable and Unspeakable in Quantum Mechanics*, 2nd ed., Cambridge University Press, 2004

[59] H. J. Bernstein, D. M. Greenberger, M. A. Horne, A. Zeilinger, "Bell theorem without inequalities for two spinless particles," Physical Review A, Vol. 47, Number 1, January, 1993

[60] J. S. Bell, Physics I, 195-200 (1964) No doubt resigned to what he took to be the unbreakable nonlocality consequences of his theorem, Bell was a proponent of Bohm's theory, although describing it as a "grossly nonlocal theory."

[61] J. Bernstein, *Quantum Profiles* (Princeton University Press), 1991, p. 82

[62] J. S. Bell, "Bertlmann's socks and the nature of reality," *Speakable and Unspeakable in Quantum Mechanics*, 2nd ed., Cambridge University Press, 2004